\documentclass[preprint,12pt]{elsarticle}

\usepackage[utf8]{inputenc}
\usepackage[T1]{fontenc}
\usepackage{amsmath,amssymb,amsthm}
\usepackage{graphicx}
\usepackage{booktabs}
\usepackage{multirow}
\usepackage{array}
\usepackage{subcaption}
\usepackage{algorithm}
\usepackage{algpseudocode}
\usepackage{listings}
\usepackage{xcolor}
\usepackage{url}
\usepackage{hyperref}
\hypersetup{colorlinks=true,linkcolor=blue,citecolor=blue,urlcolor=blue}

\definecolor{codegray}{rgb}{0.95,0.95,0.95}
\begin{document}

\begin{frontmatter}

\title{Automating structural reliability analysis with a multi-agent large
language model framework}

\author[utoronto]{Jaehwan Jeon}
\author[snu-cuee]{Chang Hee Lee} 
\author[snu-cuee,snu-ice]{Jinhyun Choo}
\author[yonsei]{Taeyong Kim\corref{cor1}}
\cortext[cor1]{Corresponding author}
\ead{tyong.kim@yonsei.ac.kr}

\affiliation[utoronto]{%
  organization={Department of Civil and Mineral Engineering, University of Toronto},
  city={Toronto},
  state={Ontario},
  country={Canada}
}
\affiliation[snu-cuee]{%
  organization={Department of Civil, Urban and Environmental Engineering, Seoul National University},
  city={Seoul},
  country={Republic of Korea}
}
\affiliation[snu-ice]{%
  organization={Institute of Construction and Environmental Engineering, Seoul National University},
  city={Seoul},
  country={Republic of Korea}
}
\affiliation[yonsei]{%
  organization={Department of Civil and Environmental Engineering, Yonsei University},
  city={Seoul},
  country={Republic of Korea}
}

\begin{abstract}
Structural reliability analysis supports the design and safety assessment of buildings and civil infrastructure but requires specialized expertise throughout the workflow. This study presents a multi-agent large language model (LLM) framework that automates component-level reliability analysis from a natural-language problem statement to interpreted estimates of the reliability index and failure probability. Specialized agents handle problem formulation, method planning, code generation, execution, and result interpretation, with human confirmation at key decision points. Numerical reliability estimates are not generated by an LLM; instead, they are obtained solely through deterministic execution of validated solvers. This design mitigates the risk of hallucinated numerical outputs and supports reproducibility. The Method Planner is fine-tuned using quantized low-rank adaptation (QLoRA) for a priori reliability-method category selection. The framework uses open-weight models and supports local execution without closed APIs. Results show that it lowers the expertise barrier to structural reliability assessment while preserving computational trustworthiness.
\end{abstract}


\end{frontmatter}

\section{Introduction}
\label{sec:introduction}

Buildings and civil infrastructure are designed and operated under uncertainty in loads, material properties, geometry, and modeling parameters. Structural reliability analysis provides a probabilistic framework for representing these uncertainties as random variables and quantifying the likelihood that a structure exceeds a prescribed limit state, commonly expressed as the probability of failure $P_f$, or the equivalent reliability index $\beta=-\Phi^{-1}(P_f)$ \citep{der2022structural}. This concept is deeply embedded in modern design codes worldwide, including partial safety factor formats and load-and-resistance-factor design, which aim to ensure adequate structural safety under uncertain demand and resistance \citep{ellingwood1980development}. In recent decades, climate change, urbanization, aging infrastructure, and increasing system complexity have further amplified the sources and consequences of uncertainty, making reliability-based assessment increasingly important for both new design and existing infrastructure management. However, conducting structural reliability analysis still requires specialized expertise in probability theory, structural mechanics, numerical simulation, and computational methods. This creates a practical barrier for the growing number of engineers and researchers who require reliability estimates but may not have advanced training in reliability theory or probabilistic computation. The present paper addresses this gap by developing an end-to-end automated reliability analysis workflow using large language models (LLMs). The proposed workflow streamlines the computational process while maintaining engineer oversight throughout the analysis.

\subsection{Background}
\label{sec:background}

\begin{figure}[t]
\centering
\includegraphics[width=0.62\linewidth]{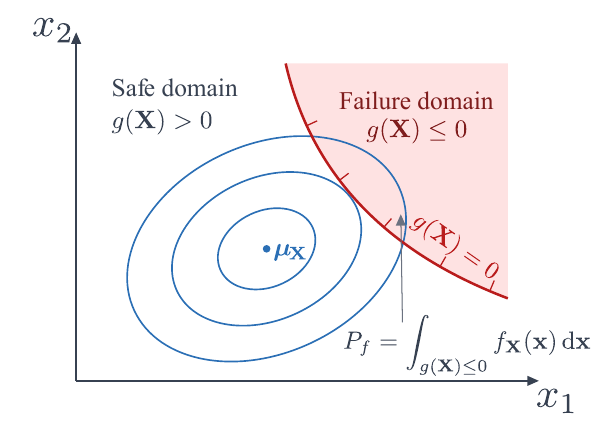}
\caption{Failure domain and the limit-state surface \(g(\mathbf{X})=0\).}
\label{fig:failure-domain}
\end{figure}

A main objective of structural reliability analysis is to estimate the probability of failure, \(P_f\), and the corresponding reliability index, \(\beta=-\Phi^{-1}(P_f)\), where \(\Phi^{-1}(\cdot)\) denotes the inverse cumulative distribution function of the standard normal distribution. In general, \(P_f\) is defined as the integral of the joint probability density function \(f_{\mathbf{X}}(\mathbf{x})\) of the random variables \(\mathbf{X}\) over the failure domain:
\begin{equation}\label{eq:pf-component}
P_f = \int_{g(\mathbf{X}) \le 0} f_{\mathbf{X}}(\mathbf{x})\, d\mathbf{x},
\end{equation}
where \(g(\mathbf{X})\) is the limit-state function, which mathematically defines the boundary between safe and failure states. Failure is assumed to occur when \(g(\mathbf{X}) \le 0\), while \(g(\mathbf{X})>0\) corresponds to the safe domain. For a two-dimensional random-variable space, the graphical relationship between the safe domain, the failure domain, and the limit-state surface \(g(\mathbf{X})=0\) is illustrated in Fig.~\ref{fig:failure-domain}. Note that this study focuses on component-level structural reliability analysis, in which failure is defined by a single limit-state function; the extension to system-level reliability involving multiple limit-state functions is deferred to future work, as discussed in Section~\ref{sec:limitations}.

A typical reliability analysis involves four major steps. The first step is \emph{problem definition}, in which the uncertain physical quantities are identified from the engineering problem and modeled as random variables \(\mathbf{X}\) with appropriate probability distributions. In this step, the limit-state function \(g(\mathbf{X})\) should be formulated to represent the failure condition of interest, such as excessive stress, displacement, ductility demand, or loss of load-carrying capacity.

The second step is \emph{method selection}. Because Eq.~\eqref{eq:pf-component} generally has no closed-form solution, an appropriate reliability method needs to be selected to evaluate or approximate the failure probability. The choice of method depends on several factors, including the dimensionality of the random-variable space, the degree of nonlinearity of the limit-state function, the presence of multiple failure regions, the target range of \(P_f\), and the computational cost of each model evaluation. Approximation methods, such as the first-order reliability method (FORM) and second-order reliability method (SORM), are widely used because they can estimate \(P_f\) efficiently by approximating the limit-state surface near the design point \citep{hasofer1974exact,rackwitz1978structural}. However, their accuracy may deteriorate when the limit-state function is highly nonlinear, when multiple design points exist, or when the failure domain has a complex geometry. Monte Carlo simulation (MCS) provides a more general and conceptually straightforward alternative because it estimates \(P_f\) directly from random samples. In principle, the MCS estimate converges to the true failure probability as the number of samples approaches infinity. However, standard MCS becomes computationally inefficient for rare-event problems, particularly when \(P_f\) is small or when each evaluation of \(g(\mathbf{X})\) requires an expensive numerical model, such as a finite-element analysis. To improve efficiency in such cases, various variance-reduction and advanced simulation techniques have been developed, including subset simulation (SuS) \citep{au2001estimation}, cross-entropy-based importance sampling \citep{geyer2019cross}, sequential importance sampling (SIS) \citep{papaioannou2016sequential}, combination line sampling (CLS) \citep{papaioannou2021combination}, and stochastic spectral embedding \citep{wagner2022rare}. However, implementing these methods remains challenging because it often requires algorithm-specific parameter selection, convergence assessment, and problem-dependent tuning; thus, their practical use can be difficult for engineers and researchers who lack specialized expertise in reliability analysis. Because each method has distinct advantages, assumptions, and limitations, selecting a suitable method is essential for achieving accurate reliability estimates with reasonable computational effort. However, applying a selection criterion may require numerical results that are unavailable before analysis, motivating the use of an AI-assisted strategy to predict the method category from the problem specification alone.

The third step is \emph{implementation}. Once a reliability method has been selected, it should be implemented consistently with the prescribed random variables, probability distributions, dependence structure, and limit-state function. In practice, this requires either developing a reliable computational implementation or correctly using existing reliability software packages, such as FERUM \citep{bourinet2010ferum} or ERA \citep{era2024software}. Errors in implementation, such as incorrect distribution parameters, inconsistent transformations of random variables, or improper convergence criteria, can significantly affect the estimated failure probability.

The fourth step is \emph{result interpretation}. The computed failure probability \(P_f\) and reliability index \(\beta\) need to be evaluated in relation to target reliability levels, design requirements, and engineering consequences. This step is not merely numerical; it requires translating probabilistic results into engineering insight, such as whether a structure satisfies a prescribed safety level, which variables dominate the failure probability, or whether additional design modifications are required.

Despite the maturity of structural reliability theory, the complete analysis workflow remains difficult for non-experts because it requires integrated knowledge of probability theory, structural mechanics, numerical methods, and computational implementation. Moreover, upstream errors or inconsistencies can propagate through the workflow, thereby compromising the final estimates of \(P_f\) and \(\beta\).

\subsection{Related work}
\label{sec:related-work}

\subsubsection{Large language models and multi-agent systems for engineering}
\label{sec:rw-llm}

Building on the demonstrated capabilities in natural-language understanding and code generation, LLMs have rapidly been applied to the automation of engineering tasks. In its simplest early form, a single LLM interprets a problem stated in natural language and directly generates and executes the analysis code. In structural and geotechnical engineering, single LLMs have been used to generate OpenSeesPy~\citep{zhu2018openseespy} code for frame analysis~\citep{liang2025integrating} and beam analysis~\citep{liu2025large}, as well as to produce structural drawings~\citep{zhang2025large} and to build geotechnical numerical models~\citep{kim2024chatgpt, kim2025can}. A single LLM, however, is prone to hallucination and error accumulation over long, multi-step tasks, producing plausible but incorrect code or numerical values~\citep{geng2026novel}. Indeed, some systems have the LLM produce numerical results directly, without executing any code~\citep{youwai2026large}, which exposes a basic limitation: the quantitative computations that reliability analysis requires cannot be entrusted to an LLM alone.

A common response to this limitation is to decompose the overall task across multiple cooperating, specialized agents and to delegate the production of numerical results to actual code execution rather than to the LLM. Such decomposition has been reported to reduce hallucination and error accumulation in multi-step operations~\citep{geng2026novel}. This pairing of a multi-agent organization with code execution has become established across civil and construction engineering, with examples in reinforced-concrete design~\citep{chen2025multi}, BIM coordination~\citep{dong2025ai}, structural design workflows~\citep{liang2025automating}, frame modeling~\citep{geng2025lightweight, geng2026agentic}, geotechnical design~\citep{xu2025multimodal}, and blast-safety assessment~\citep{wang2026towards}, as summarized in Table~\ref{tab:llm-engineering}. A multi-agent organization combined with deterministic code execution has thus become a widely adopted, common practice.

\begin{table}[!tp]
  \centering
  \fontsize{7.5}{8.5}\selectfont
  \setlength{\tabcolsep}{2.5pt}
  \renewcommand{\arraystretch}{0.92}
  \caption{Application of large language models in engineering. None of the
  surveyed works targets structural reliability analysis ($\beta$/$P_f$)
  itself or fine-tunes the \emph{a priori} selection of the analysis method.}
  \label{tab:llm-engineering}
  \begin{tabular}{@{}l p{1.45cm} p{1.45cm} p{1.35cm} p{1.0cm} p{1.6cm} p{1.85cm}@{}}
    \toprule
    Reference & Domain & Task & LLM(s) & Agents & Fine-tuning (target) & Executes? \\
    \midrule
    \citet{chen2025multi}        & RC design           & Code-compliant design                         & DeepSeek-V3, GPT-4o         & Multi (4)      & No          & Yes (Python)     \\
    \citet{dong2025ai}           & BIM                 & NL BIM coordination                           & o1-mini, GPT-4              & Multi (5)      & No          & Yes (Revit API)  \\
    \citet{liang2025integrating} & Structural analysis & NL frame analysis                             & GPT-4o, Gemini, Llama       & Single         & No          & Yes (OpenSeesPy) \\
    \citet{liang2025automating}  & Structural design   & Design workflows                              & GPT-4o, Claude              & Multi          & No          & Yes (OpenSeesPy) \\
    \citet{geng2025lightweight}  & Structural analysis & 2D frame modeling                             & Llama-3.3 70B               & Multi (5)      & No          & Yes (OpenSeesPy) \\
    \citet{geng2026agentic}      & Structural analysis & 3D frame modeling                             & GPT-OSS, Llama-3.3          & Multi          & No          & Yes (SAP2000)    \\
    \citet{zhang2025large}       & Structural drawings & NL to drawings                                & ChatGPT-3.5/4               & Single         & No          & Yes (AutoCAD)    \\
    \citet{liu2025large}         & Structural analysis & Beam analysis                                 & Llama-3.3 70B               & Single         & No          & Yes (OpenSeesPy) \\
    \citet{youwai2026large}      & Foundation design   & Foundation design calc.                       & DeepSeek, GPT, Grok, Gemini & Multi (router) & No          & No               \\
    \citet{kim2024chatgpt}       & Geo\-technical      & Numerical modeling                            & ChatGPT-4                   & Single         & No          & Yes (MATLAB)     \\
    \citet{kim2025can}           & Geo\-technical      & FE modeling                                   & ChatGPT o1                  & Single         & No          & Yes (FEniCS)     \\
    \citet{xu2025multimodal}     & Geo\-technical      & Footing design                                & GPT-4o                      & Multi (3)      & No          & Yes (Python)     \\
    \citet{wang2026towards}      & Blast safety        & Safety assessment                             & Qwen2.5, DeepSeek-R1        & Multi (4)      & No          & Yes (Python)     \\
    \citet{shi2025fine}          & Building codes      & Compliance scripting                          & Mistral 7B                  & Single         & QLoRA -- gen.            & No               \\
    \citet{jiang2025efficient}   & Building energy     & NL to energy model                            & T5-11B                      & Single         & LoRA -- gen.             & Yes (EnergyPlus) \\
    \citet{dong2025fine}         & CFD                 & NL to CFD setup                               & Qwen2.5-7B                  & Multi (4)      & LoRA -- gen.             & Yes (OpenFOAM)   \\
    \citet{zhang2026automating}  & Contracts           & Contract QA                                   & DeepSeek-R1-Qwen-14B        & Single         & SFT+RL -- QA             & No               \\
    \citet{aqib2025fine}         & Building codes      & Code QA (RAG)                                 & Qwen, Llama, Phi            & Single         & PEFT -- QA               & No               \\
    \midrule
    \textbf{This work}           & \textbf{Structural reliability} & Reliability analysis ($\beta$/$P_f$) + method selection & Gemma~4 (E4B + 26B-A4B) & Multi (5) & \textbf{QLoRA -- \emph{a priori} sel.} & Yes (det.) \\
    \bottomrule
  \end{tabular}

  \smallskip
  \parbox{\linewidth}{\raggedright\scriptsize Note: In the Gemma~4 model names, E4B denotes nominally 4B (billion) effective parameters, and 26B-A4B denotes a mixture-of-experts model~\citep{shazeer2017outrageously} with nominally 26B total and 4B active parameters. Abbreviations: NL, natural language; PEFT, parameter-efficient fine-tuning; SFT, supervised fine-tuning; RL, reinforcement learning; gen., generation; sel., selection; det., deterministic.}
\end{table}

\subsubsection{Domain adaptation and fine-tuning of LLMs}
\label{sec:rw-finetuning}

Although general-purpose LLMs often perform well on a wide range of tasks, their performance can degrade when they are applied to domain-specific and technically complex problems. To address this limitation, two broad adaptation strategies are commonly used. The first strategy keeps the model weights fixed and improves the input context through techniques such as prompt engineering or retrieval-augmented generation (RAG), in which relevant external knowledge is retrieved and appended to the user query. The second strategy modifies the model weights through fine-tuning, thereby enabling deeper domain specialization. Among fine-tuning approaches, QLoRA \citep{dettmers2023qlora} has become particularly attractive because it combines low-rank adaptation (LoRA) \citep{hu2022lora} with 4-bit quantization, allowing large models to be fine-tuned efficiently on a single consumer-grade GPU.

Fine-tuning has recently begun to appear in civil and construction engineering applications, mainly in two task categories. The first category is generation. For example, \citet{shi2025fine} fine-tuned a model using QLoRA to generate building-code-compliance scripts, while \citet{jiang2025efficient} fine-tuned a model to generate building-energy models from natural-language descriptions. A similar pattern is observed in multi-agent workflows that execute analysis code, where fine-tuning is primarily used to translate problem descriptions into solver configurations \citep{dong2025fine}. The second category is question answering. \citet{zhang2026automating} applied supervised fine-tuning with reinforcement learning to construction-contract question answering, and \citet{aqib2025fine} evaluated fine-tuning together with retrieval for building-code question answering.

\subsection{Knowledge gaps}
\label{sec:knowledge-gaps}

Structural reliability analysis requires an integrated workflow that connects problem formulation, method selection, numerical computation, and result interpretation. However, as summarized in Table~\ref{tab:llm-engineering}, this workflow has not yet been systematically addressed in existing LLM-based engineering studies.

This review reveals two major knowledge gaps. First, no existing LLM-based framework provides end-to-end automation for structural reliability analysis, from natural-language problem formulation to the estimation and interpretation of \(P_f\) and \(\beta\). Existing reliability tools require expert-defined inputs, whereas general-purpose LLMs cannot be directly trusted for quantitative reliability analysis because of hallucination and the need for deterministic computation.

Second, and more importantly, the \emph{a priori} selection of an appropriate reliability-analysis method remains largely unexplored in LLM-based engineering research. Method selection is not merely a code-generation or solver-routing task; its difficulty lies in predicting the appropriate method category before the numerical results required by a selection criterion are available, rather than in applying that criterion once those results are known. The analyst needs to infer, from the problem definition alone, which method is likely to provide sufficient accuracy at a reasonable computational cost. This decision requires simultaneous consideration of the dimensionality of the random-variable space, the nonlinearity and geometry of the limit-state surface, the possible presence of multiple failure regions, the expected rarity of failure, and the cost of each limit-state evaluation. An unsuitable choice can produce either excessive computational demand or, more critically, a numerically valid but inaccurate estimate of \(P_f\) and \(\beta\).

Existing fine-tuning studies in civil and construction engineering have primarily targeted generation tasks, such as producing analysis scripts or solver configurations, or question-answering tasks based on technical documents. These tasks do not address the distinct challenge of learning to predict, from a problem specification alone, the method category assigned by a reference selection criterion within an executable reliability-analysis pipeline. Thus, this gap motivates investigating whether domain-specific fine-tuning of an LLM component improves \emph{a priori} method-category selection from a structured problem description within a locally executable, open-weight framework, without first executing all candidate methods or having access to the unknown reference solution.

\subsection{Contributions}
\label{sec:contributions}

This work proposes an integrated, locally deployable multi-agent LLM framework for structural reliability analysis. Its primary contribution is a workflow that carries a component-level reliability problem from a natural-language statement to interpreted estimates of the reliability index and failure probability, in which the user confirms or corrects key intermediate decisions. The central design principle of this workflow is the separation of language-based reasoning from numerical computation. The LLM agents formulate the problem, select the method category, generate the solver code, and explain the result. All numerical estimates are obtained from deterministic solver execution, which mitigates the risk of hallucinated numerical outputs.

This workflow is enabled by three design features: (i) decomposition into specialized agents connected by typed data contracts, which makes each intermediate output checkable and traceable; (ii) human-in-the-loop confirmation and correction at decision points that affect the course of the analysis; and (iii) the exclusive use of open-weight models, which allows local execution without dependence on closed APIs. As a component-level specialization within this workflow, only the Method Planner is fine-tuned using QLoRA, with the aim of improving the accuracy of \emph{a priori} method-category selection while keeping the adapted Planner within the same locally executable, open-weight workflow.

\subsection{Paper organization}
\label{sec:organization}

Section~\ref{sec:approach} presents the proposed multi-agent framework: the problem formulation, the structured data contracts and agent design, the QLoRA fine-tuning of the Method Planner, the user interface, and an end-to-end demonstration on a running example. Section~\ref{sec:experiments} evaluates the integrated workflow on twenty benchmark problems in terms of code execution and numerical accuracy, including a comparison with direct single-LLM answers, and separately compares the method-category selection of the fine-tuned Planner with the untuned configurations. Section~\ref{sec:discussion} discusses the implications and limitations of the results, and Section~\ref{sec:conclusion} concludes the paper.

\section{Proposed approach}
\label{sec:approach}

This section presents the architecture of the multi-agent framework and the design of the individual agents (Section~\ref{sec:overview}), the fine-tuning of the Method Planner (Section~\ref{sec:finetuning}), the user interface (Section~\ref{sec:ui}), and an end-to-end walkthrough of the running example---a slender steel column under Euler buckling (Section~\ref{sec:running-example}).

\subsection{Multi-agent framework}
\label{sec:overview}

\begin{figure}[t]
  \centering
  \includegraphics[width=\linewidth]{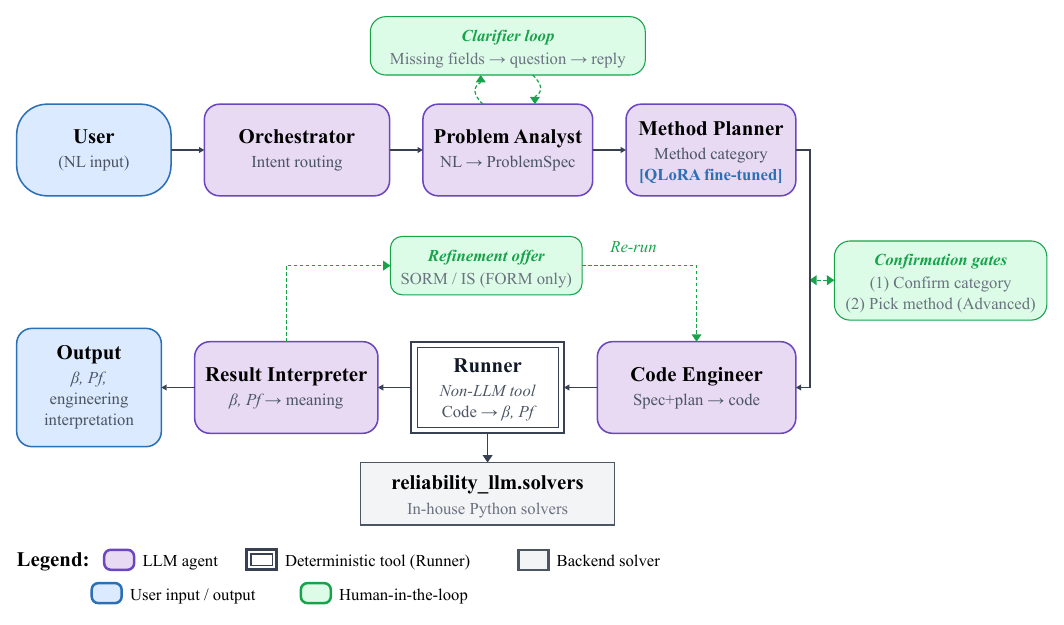}
  \caption{Multi-agent framework for structural reliability analysis. Green
    marks the human-in-the-loop interaction points. At any point, the Orchestrator can answer general questions directly or route to the Method Planner or Result Interpreter (paths not shown).}
  \label{fig:architecture}
\end{figure}

The proposed framework consists of five LLM-based agents and one deterministic execution component, which are sequentially connected as illustrated in Fig.~\ref{fig:architecture}: the Orchestrator, Problem Analyst, Method Planner, Code Engineer, Runner, and Result Interpreter. The \textbf{Orchestrator} classifies user intent and routes it to the appropriate agent, or answers general questions directly. The \textbf{Problem Analyst} converts natural-language input into a structured problem specification---the random variables with their distributions, the limit-state function, and the correlation structure. The \textbf{Method Planner} selects the method category suited to the problem; it is the only fine-tuned agent, specialized to recommend the appropriate category (Section~\ref{sec:finetuning}). The \textbf{Code Engineer} turns the problem specification and the selected method into executable code that calls the backend solver. The \textbf{Runner} is a deterministic, non-LLM tool that executes this code in an isolated subprocess and extracts the reliability index $\beta$ and the failure probability $P_f$. Lastly, the \textbf{Result Interpreter} turns these numerical results into an engineering interpretation.

In the proposed framework, three aspects are particularly important. First, the intermediate outputs of the reliability-analysis workflow are produced sequentially by the corresponding agents. The LLM agents do not autonomously revise or repeat the outputs of previous stages, and the interfaces between stages are fixed by the data contracts defined in Section~\ref{sec:contracts}. Second, the reliability index \(\beta\) and the failure probability \(P_f\) are never generated directly by an LLM. They are produced solely by the Runner, which is a deterministic non-LLM component that executes the solver code. The Runner parses \(\beta\) and \(P_f\) from the output of the executed code and passes them to the Result Interpreter through the \texttt{ExecutionResult} contract. If execution fails, the Runner returns structured error information instead, and the Result Interpreter reports the failure rather than interpreting unavailable results. This design makes the numerical results reproducible and verifiable, while reducing the risk that LLM hallucinations propagate into the reported reliability estimates. Third, this sequential ordering applies only to the internal analysis workflow; the user-facing interaction is not constrained to a fixed linear sequence, as described in Section~\ref{sec:confirmation}. 

\subsubsection{Structured data contracts}
\label{sec:contracts}

The proposed framework is designed to follow the logical structure of structural reliability analysis. Each hand-off between agents is represented by a typed object, hereafter referred to as a \emph{data contract} following the design-by-contract principle \citep{meyer1992applying}. Each data contract specifies the required fields, their formats, and the information to be passed to the next stage. An agent's output is forwarded only after it satisfies the corresponding specification. 

By explicitly encoding the domain structure in the aforementioned way, errors such as missing random variables, malformed fields, inconsistent distribution definitions, or ill-formed limit-state functions can be detected at the stage where they are produced. Furthermore, because all intermediate outputs are stored in structured form, the analysis remains traceable: the reported reliability index \(\beta\) and failure probability \(P_f\) can be traced back to the problem specification, selected method, generated code, and execution result. This traceability supports the reproducibility required for reliability assessment in design-verification contexts.

From the user's natural-language input to the final engineering interpretation, the framework defines five core data contracts: \texttt{ProblemSpec} for problem definition, \texttt{AnalysisPlan} for method-category selection, \texttt{GeneratedCode} for executable code, \texttt{ExecutionResult} for the computed \(\beta\) and \(P_f\), and \texttt{EngineeringInterpretation} for the result summary, reliability-level assessment, target comparison, and caveats. Table~\ref{tab:contracts} summarizes these data contracts and the information passed between stages. The routing and confirmation logic is described separately in Section~\ref{sec:orchestration} and Section~\ref{sec:confirmation}.

\begin{table}[!htb]
  \centering
  \caption{The five core data contracts, the agent that produces each, and the
    information each passes to the next stage.}
  \label{tab:contracts}
  \small
  \begin{tabular}{@{}l l p{0.5\linewidth}@{}}
    \toprule
    Data contract & Produced by & Information passed downstream \\
    \midrule
    \texttt{ProblemSpec} & Problem Analyst & Random variables (each with its
      distribution, mean, and standard deviation), the limit-state function
      $g(\mathbf{X})$, and the correlation matrix \\
    \texttt{AnalysisPlan} & Method Planner & A single recommended method
      category (FORM-based, MCS, or Advanced) \\
    \texttt{GeneratedCode} & Code Engineer & Executable Python code that calls
      the backend solver \\
    \texttt{ExecutionResult} & Runner & Whether execution succeeded, with the
      reliability index $\beta$ and failure probability $P_f$ (or error
      information on failure) \\
    \begin{tabular}[t]{@{}l@{}}\texttt{Engineering}\\\texttt{~~Interpretation}\end{tabular}
      & Result Interpreter & A result summary, a
      reliability level, a comparison to the target, and caveats \\
    \bottomrule
  \end{tabular}
\end{table}

\subsubsection{Agent design}
\label{sec:agents}

This section presents the internal design of each agent, focusing on the distinctive mechanism by which each component processes its input and passes structured information to the next stage.

\paragraph{Orchestrator}
\label{sec:agent-orchestrator}
The Orchestrator serves as the entry point of the framework and classifies each incoming user message into one of five intents: a new problem, a modification to an existing problem, a request to explain a result, a general question, or a reply to a clarification request. Although this step does not involve reliability computation, it determines the downstream path of the workflow and therefore functions as the system's gateway. By representing the user intent as an explicit type, the Orchestrator separates routing from domain-specific reasoning, allowing subsequent tasks such as problem extraction, method selection, code generation, and interpretation to be handled by dedicated agents.

\paragraph{Problem Analyst}
\label{sec:agent-analyst}
The Problem Analyst converts free-form natural-language input into a structured problem specification, represented by the \texttt{ProblemSpec} data contract. It extracts the random variables, their probability distributions and statistical parameters, the limit-state function, and the correlation structure. It also maps loosely stated distribution names in the user input to the distribution families supported by the backend solver. If a required item is missing or ambiguous, the Problem Analyst does not infer the missing information by assumption; instead, it initiates the clarification loop described in Section~\ref{sec:confirmation}.

\paragraph{Method Planner}
\label{sec:agent-planner}
The Method Planner (hereafter the Planner) receives the \texttt{ProblemSpec} and selects a single reliability-method category suited to the problem. Among the various methods available for structural reliability analysis, this study considers three categories: (i) the FORM-based category, which includes FORM, SORM, and FORM-based importance sampling; (ii) the MCS category, which corresponds to crude Monte Carlo simulation; and (iii) the Advanced category, which includes advanced sampling techniques such as subset simulation, cross-entropy-based importance sampling, sequential importance sampling, and combination line sampling. The selection of the specific method within the chosen category is deferred to a later stage, as described in Section~\ref{sec:confirmation}. 

Selecting an appropriate category \emph{a priori} is particularly challenging because the decision needs to be made before any reliability analysis has been performed and before the resulting \(P_f\), \(\beta\), convergence behavior, or method-specific approximation error is known. The Planner should therefore infer, from the problem specification alone, which category is likely to provide sufficient accuracy at a reasonable computational cost. This inference requires simultaneous consideration of problem characteristics such as dimensionality, limit-state nonlinearity, failure-domain geometry, expected failure rarity, and model-evaluation cost. In practice, the most reliable way to identify the best-performing method would be to execute multiple candidate methods and compare their results; however, doing so would defeat the purpose of efficient method selection. Because this difficult pre-analysis decision requires substantial domain-specific judgment, the Planner is the only fine-tuned agent in the framework. The fine-tuning procedure and rationale are described in Section~\ref{sec:finetuning}. Although the present study focuses on these three categories, the same procedure can be extended to additional reliability-analysis methods in future work.

\paragraph{Code Engineer}
\label{sec:agent-code-engineer}
The Code Engineer receives the problem specification and the selected method, and generates complete, executable Python code that calls the backend reliability solver. To support this process, we developed a unified solver library based on
established reliability-analysis literature and software \citep{der2022structural,bourinet2010ferum,era2024software}. Moreover, the generated code is required to satisfy a fixed output format: it prints \(\beta\) and \(P_f\) on the final two lines in a predefined format so that the Runner can parse the results deterministically. Thus, the Code Engineer does not generate arbitrary free-form analysis code; instead, it translates the structured specification into a standardized solver call with a fixed interface to the execution stage.

\paragraph{Runner}
\label{sec:agent-runner}
The Runner is a deterministic, non-LLM execution component that operates as a node in the workflow graph. It executes the code generated by the Code Engineer in an isolated subprocess and parses \(\beta\) and \(P_f\) from the standard output using the predefined format. If both values are successfully parsed, the execution is marked as successful. Otherwise, the Runner returns structured error information, including the failure type, such as a parse failure or runtime exception. Executing the generated code in isolation prevents faults in the code from affecting the wider system, while ensuring that the subsequent interpretation stage always receives a structured execution result.

\paragraph{Result Interpreter}
\label{sec:agent-interpreter}
The Result Interpreter converts the numerical outputs of the Runner into an engineering interpretation. It summarizes the reliability result, classifies the reliability level into one of five discrete categories ranging from very low to very high, compares the result with the target reliability level, and reports relevant caveats. If the execution fails, the Result Interpreter reports the failure and its type rather than attempting to interpret unavailable or invalid numerical results.

Among the six components, the full system prompt of the Planner is reproduced in \ref{app:planner-prompt}, because the fine-tuning procedure in Section~\ref{sec:finetuning} and the evaluation in Section~\ref{sec:experiments} depend directly on this prompt.

\subsubsection{Framework flow}
\label{sec:orchestration}

Each user message is first received by the Orchestrator. When the
message describes a new reliability problem, the Orchestrator passes it
to the Problem Analyst to start the analysis, and the workflow then
proceeds serially in the order shown in Fig.~\ref{fig:architecture}: the
Planner recommends a method category, the Code Engineer generates the
solver-calling code, the Runner executes it, and the Result Interpreter
translates the numerical results into an engineering interpretation.
General questions that do not require a reliability analysis are
answered directly without entering the pipeline.

When the session already contains a completed analysis, the Orchestrator
instead routes the message to an intermediate point of the pipeline.
Requests that modify the problem specification return to the Problem
Analyst, which updates the specification; requests to repeat the
analysis with a different method are sent to the Planner; and requests
for further explanation of a result are sent to the Result Interpreter.
In each case, the workflow resumes from the routed point and proceeds in
the same order as before. Whether on a first run or on re-entry, the
progression pauses at four points where additional information or user
confirmation is required, as described in Section~\ref{sec:confirmation}.

\subsubsection{Human-in-the-loop interaction}
\label{sec:confirmation}

The proposed framework does not execute the full reliability-analysis workflow as an unattended black box. Instead, it adopts a human-in-the-loop design \citep{wu2022survey}, in which routine computational steps proceed automatically but the workflow pauses at decision points that can affect the course of the analysis. This allows non-expert users to retain control over important choices, while the agents handle the technical tasks of problem structuring, method planning, code generation, execution, and result interpretation. 

The workflow pauses for user input at four points (Table~\ref{tab:routing}). First, during problem definition, the Problem Analyst may identify missing or ambiguous items in the extracted \texttt{ProblemSpec}. In that case, the Clarifier requests only the required information from the user, asking again if necessary, and the workflow resumes once the replies complete the problem specification. This prevents incomplete or ambiguous inputs from propagating into method selection and numerical analysis. Second, after the Planner recommends a reliability-method category, the system asks the user to confirm the recommendation or switch to another category. This gate allows the user to retain control over the overall analysis strategy before code generation and numerical execution begin. Third, if the Advanced category is confirmed, the user is asked to select a specific advanced sampling method before code generation begins. This additional gate is required because the Advanced category includes multiple methods with different assumptions, tuning requirements, and computational characteristics. Fourth, after a first-pass FORM-based analysis is completed, the system offers a one-time refinement using SORM or FORM-based importance sampling when a more accurate or robust estimate may be desirable.

At each gate, the user can either proceed with the proposed step, request an explanation, or redirect the conversation. When an explanation is requested, the system answers the question and then returns to the same decision point. When the user changes to an unrelated topic, the message is routed back to the Orchestrator. This design preserves the sequential structure of the internal analysis workflow while allowing flexible user-facing interaction.

\begin{figure}[!tp]
  \centering
  \includegraphics[width=\linewidth]{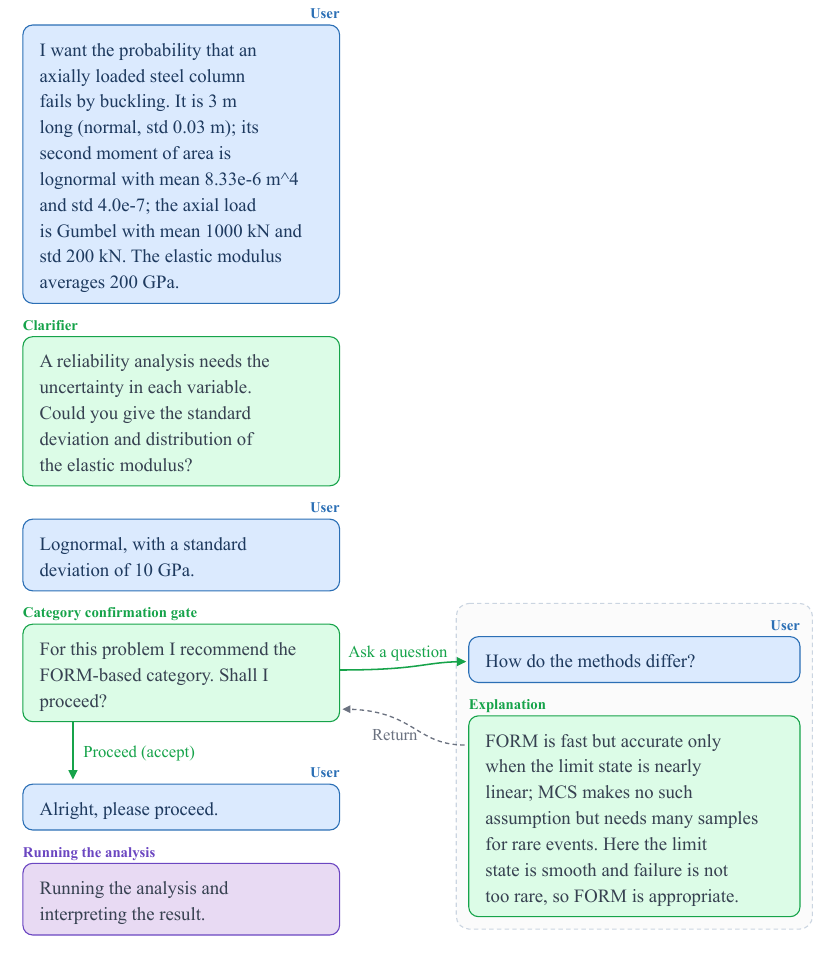}
  \caption{A representative human-in-the-loop interaction for the running-example
    column problem (Section~\ref{sec:running-example}). User messages are shown on
    the right and system messages on the left; green marks the
    human-in-the-loop interaction points---the Clarifier requesting a missing
    input, the category-confirmation gate, and an on-demand explanation that
    returns to the gate (curved arrow). At each gate the user may accept, ask a
    question, or change direction, allowing a non-expert to steer the analysis
    without expert knowledge.}
  \label{fig:hitl}
\end{figure}

Fig.~\ref{fig:hitl} illustrates a representative human-in-the-loop interaction using the running-example column problem. In this example, the Clarifier first requests a missing input, the user then asks for an explanation at the category-confirmation gate, and the analysis proceeds after the user confirms the recommended category. These confirmation stages allow non-expert users to guide the analysis at key decision points without needing to understand the full internal machinery of structural reliability analysis.

\begin{table}[!tp]
  \centering
  \caption{The four human-in-the-loop gates at which the workflow pauses
    for user input. At any gate, a question is answered and the same gate
    is presented again, and a message unrelated to the pending decision
    is routed back to the Orchestrator; declining the refinement offer
    ends the analysis.}
  \label{tab:routing}
  \small
  \begin{tabular}{@{}l p{0.30\linewidth} p{0.34\linewidth}@{}}
    \toprule
    Gate & User reply & Next step \\
    \midrule
    Clarifier & Supplies the missing information & Problem Analyst
      (specification completed) \\
    Category confirmation & Accepts or switches the category & Code
      Engineer for FORM-based and MCS (automatic); the gate below for
      Advanced \\
    Advanced method selection & Picks a specific method (e.g., SuS, SIS,
      CLS) & Code Engineer \\
    Refinement offer & Accepts SORM or FORM-based importance sampling &
      Code Engineer (re-run) \\
    \bottomrule
  \end{tabular}
\end{table}

\subsection{Fine-tuning the Planner}
\label{sec:finetuning}

In the proposed framework, all LLM-based agents are used without task-specific fine-tuning, except for the Planner. This design choice represents the distinct role of the Planner in the workflow. While the other agents mainly perform language understanding, structured extraction, code generation, or result explanation under task-specific prompts, the Planner makes an \emph{a priori} decision about which reliability-method category is most appropriate for a given problem. In this study, the Planner selects one of three categories: FORM-based, MCS, or Advanced. This decision is difficult because the most suitable method is the one that achieves sufficient accuracy with reasonable computational cost, but this trade-off is generally not known before the analysis is performed. As discussed in Section~\ref{sec:background}, each category has distinct advantages and limitations. Selecting an appropriate category therefore requires expert judgment that simultaneously considers the dimensionality of the problem, the nonlinearity of the limit-state function, the expected magnitude of the failure probability, and the computational cost of model evaluations. 

\begin{figure}[!tb]
  \centering
  \includegraphics[width=0.85\linewidth]{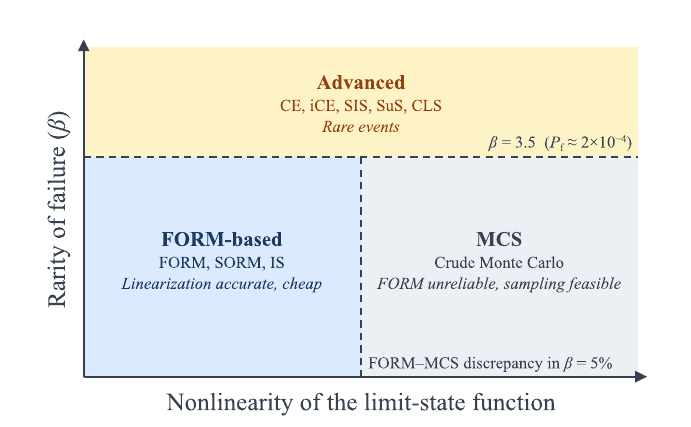}
  \caption{The deterministic reference criterion used to assign training labels. Rare failures ($\beta > 3.5$) route to Advanced regardless of nonlinearity (rarity takes precedence); otherwise the label is FORM-based when FORM's linearization stays within 5\% of the MCS reliability index, and MCS when it does not. The boundary values ($\beta = 3.5$, 5\%) are a replaceable design choice.}
  \label{fig:selection-map}
\end{figure}

To embed this selection expertise into the Planner, this study fine-tunes the Planner using QLoRA~\citep{dettmers2023qlora,hu2022lora}, with the aim of improving the accuracy of \emph{a priori} method-category selection within the locally executable, open-weight framework. The training objective is to map a structured problem specification to the method category that would be acceptable to an expert prior to conducting the analysis. To construct the training labels, we define a deterministic reference standard. Each training problem is solved offline using all candidate solvers in the backend library, and the resulting reference estimates are compared to assign the method-category label. In particular, rare-event problems with \(\beta > 3.5\) are assigned to the Advanced category, regardless of the degree of nonlinearity, because rarity takes precedence in the method-selection criterion. For the remaining problems, the label is assigned to the FORM-based category when the FORM estimate remains within 5\% of the MCS reference reliability index; otherwise, the problem is assigned to the MCS category. The threshold \(\beta=3.5\) is adopted as a representative target reliability level commonly used in design-oriented reliability assessment. Fig.~\ref{fig:selection-map} illustrates how this criterion partitions the rarity--nonlinearity plane into the three method-category regions. The labeling rule used in this study represents one possible expert-defined criterion. If a different criterion is preferred, the training labels can be regenerated accordingly, and the Planner can be retrained using the same QLoRA-based procedure.

The Planner is fine-tuned on 150 synthetic supervised examples, each mapping a problem specification to a method category. Each label is assigned using the same deterministic reference standard later used for the held-out benchmark in Section~\ref{sec:benchmark}. The model input is constructed from the problem specification alone, using the same input format during fine-tuning and deployment. Fine-tuning uses QLoRA with Gemma~4 E4B~\citep{gemma2026technical} as the base model. The base weights are quantized to 4 bits and frozen, and only a low-rank adapter is trained with rank 16 and \(\alpha=16\). The model is trained for 4 epochs using a learning rate of \(2\times10^{-4}\), an effective batch size of 4, the 8-bit AdamW optimizer~\citep{loshchilov2019decoupled}, and a linear learning-rate schedule. Training and subsequent evaluation in Section~\ref{sec:experiments} are conducted on a single RTX~A5000 GPU with 24~GB of memory. The implementation uses Unsloth's \texttt{FastModel} loader~\citep{unsloth2023software}, with only the language layers trained and the vision tower kept frozen, allowing the 4-bit QLoRA setup to fit within the available GPU memory.

\begin{figure}[!tb]
  \centering
  \includegraphics[width=\linewidth]{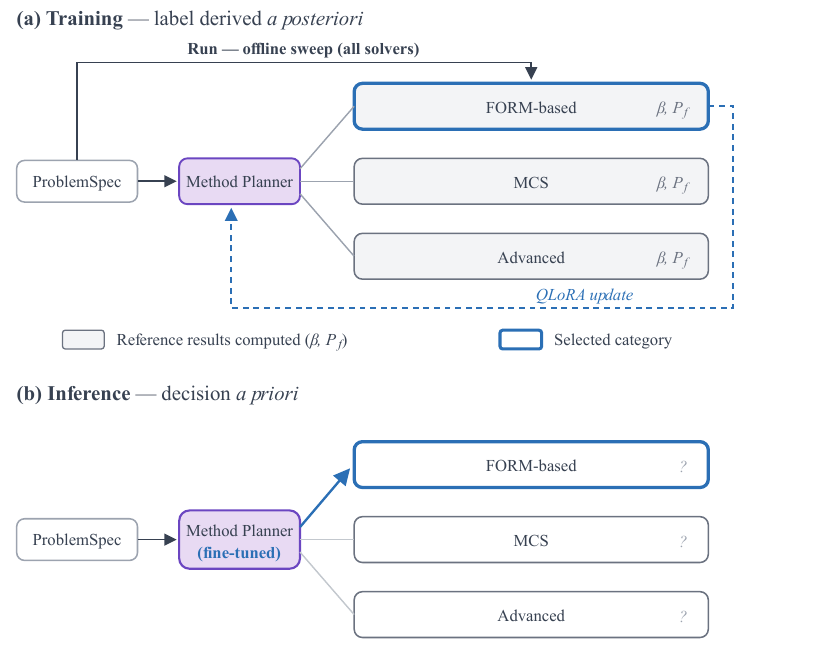}
  \caption{Training versus inference for the Planner. (a)~During training, each problem is solved once, offline, with every solver in the backend library; the gold category assigned by the deterministic rule supervises the QLoRA update. (b)~At deployment, no reference results exist: the fine-tuned Planner selects the category \emph{a priori}, from the problem specification alone. Gray-shaded boxes carry the computed reference results ($\beta$, $P_f$); the blue border marks the selected category.}
  \label{fig:train-vs-infer}
\end{figure}

The key distinction between training and deployment is that the method-category labels are obtained from \emph{a posteriori} information during training but should be predicted \emph{a priori} during inference. During training, the costly reference computations are performed only once offline to generate the supervised labels. At deployment, however, such reference results are unavailable because obtaining \(\beta\) and \(P_f\) is precisely the purpose of the reliability analysis. The fine-tuned Planner is therefore trained to predict the category assigned by this reference criterion from the problem specification alone, without running the candidate methods at inference. This distinction is illustrated in Fig.~\ref{fig:train-vs-infer}: during training, the reference results are available for label generation, whereas during inference, the Planner selects the method category before numerical computation is performed.

\subsection{User interface}
\label{sec:ui}

\begin{figure}[p]
  \centering
  \includegraphics[width=\linewidth]{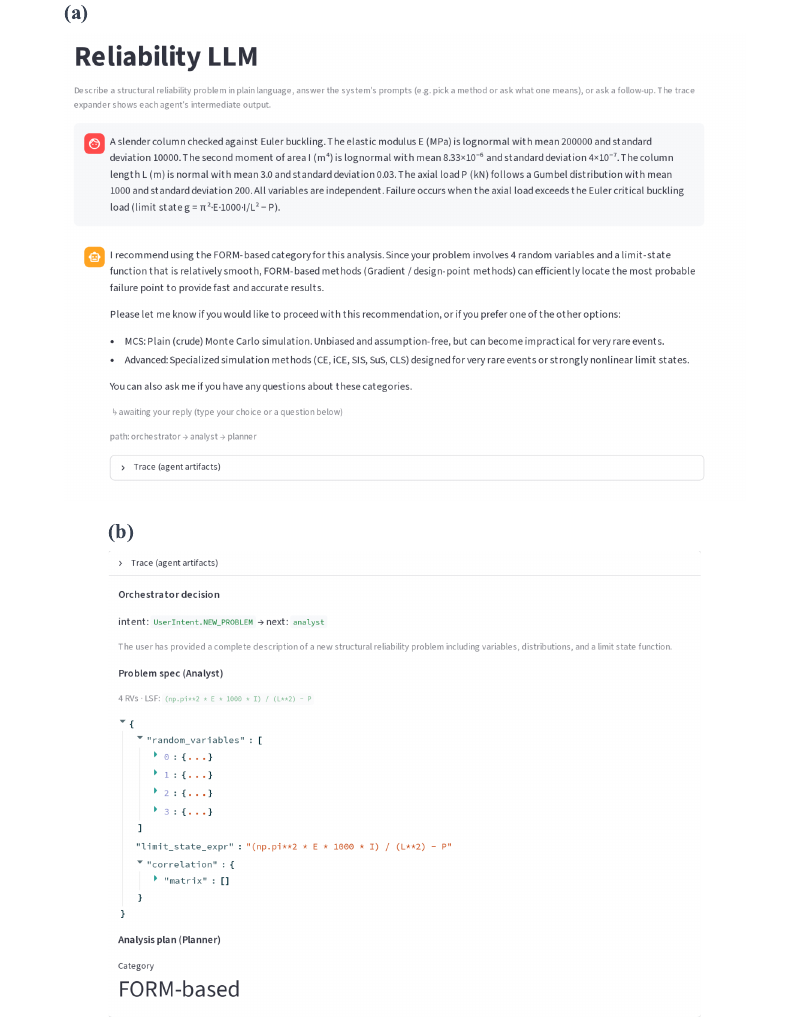}
  \caption{Chat-based web interface of the framework, shown for the
    running-example column problem (Section~\ref{sec:running-example}):
    (a)~the natural-language input and the category-confirmation gate;
    (b)~the trace panel of the same turn, exposing the data contracts of
    Table~\ref{tab:contracts}.}
  \label{fig:gui}
\end{figure}

To make the preceding workflow accessible to non-expert users, the framework provides a chat-based web interface implemented with Streamlit \citep{streamlit2019software}, which runs locally together with the agents and the backend solvers. The user describes a reliability problem in natural language, and the system replies with the outputs of the workflow stages described in Section~\ref{sec:orchestration}. The human-in-the-loop gates appear as explicit pauses within this conversation: when the workflow reaches a decision point, the interface indicates that the system is awaiting a reply, and the user responds through the same chat input, as summarized in Table~\ref{tab:routing}. Fig.~\ref{fig:gui}(a) shows this interaction for the running example (Section~\ref{sec:running-example}), where the category recommended by the Planner is presented at the confirmation gate.

In addition to the conversational replies, each system turn provides a trace panel that exposes the intermediate outputs of that turn, which correspond one-to-one to the data contracts of Table~\ref{tab:contracts}. Fig.~\ref{fig:gui}(b) shows this panel after problem extraction and method planning. The trace panel thus makes the traceability provided by the data contracts (Section~\ref{sec:contracts}) directly accessible in the interface: the user can inspect how the reported $\beta$ and $P_f$ were obtained without leaving the conversation.

\begin{figure}[!tb]
  \centering
  \includegraphics[width=\linewidth]{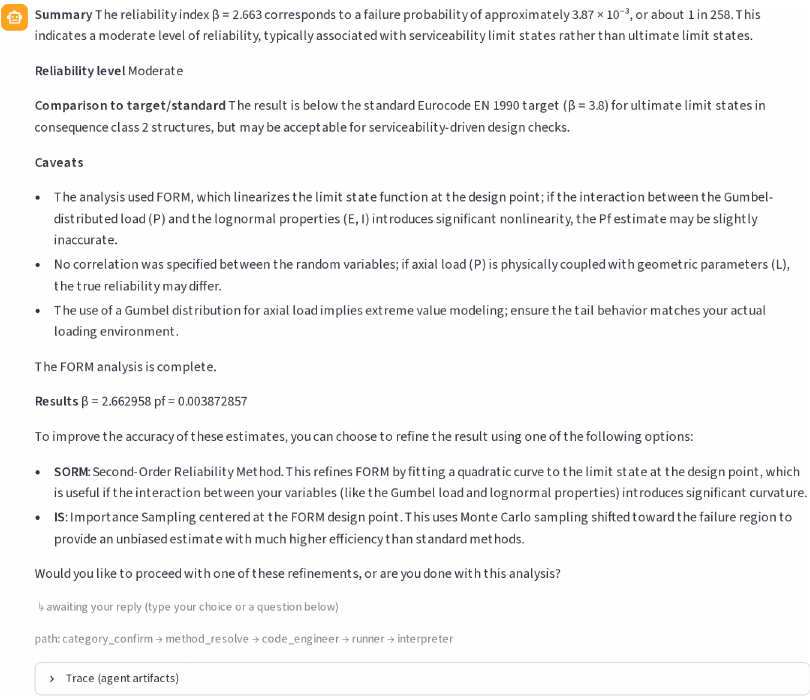}
  \caption{Closing reply of the analysis turn for the running example:
    the engineering interpretation of the computed $\beta$ and $P_f$,
    followed by the one-time refinement offer
    (Section~\ref{sec:confirmation}).}
  \label{fig:gui-result}
\end{figure}

The analysis turn closes with a conversational reply that returns the interpreted result to the user. Fig.~\ref{fig:gui-result} shows this closing reply for the running example: the engineering interpretation of the computed $\beta$ and $P_f$ is followed by the one-time refinement offer available after a first-pass FORM-based analysis (Section~\ref{sec:confirmation}).

\subsection{Running example}
\label{sec:running-example}

This section demonstrates the proposed multi-agent framework using the buckling reliability of a slender steel column under axial compression. The user provides the following natural-language input.

\begin{quote}
``A slender column checked against Euler buckling. The elastic modulus $E$ (MPa) is lognormal with mean 200000 and standard deviation 10000. The second moment of area $I$ (m\textsuperscript{4}) is lognormal with mean $8.33\times10^{-6}$ and standard deviation $4\times10^{-7}$. The column length $L$ (m) is normal with mean 3.0 and standard deviation 0.03. The axial load $P$ (kN) follows a Gumbel distribution with mean 1000 and standard deviation 200. All variables are independent. Failure occurs when the axial load exceeds the Euler critical buckling load (limit state $g = \pi^2 \cdot E \cdot 1000 \cdot I / L^{2} - P$).''
\end{quote}

The limit-state function is the Euler-buckling margin $g(\mathbf{X}) = P_{cr} - P$, where  $P_{cr} = \pi^2 E I / L^2$ is the critical buckling load. The factor of $10^{3}$ in the quoted expression converts $E$ from MPa to kN\,m$^{-2}$, so that $P_{cr}$ and $P$ are expressed in the same unit, kN. Failure (\(g(\mathbf{X}) \le 0\)) occurs when the applied axial load exceeds the buckling capacity. The four random variables are summarized in Table~\ref{tab:running-example-rvs}.

\begin{table}[t]
  \centering
  \footnotesize
  \setlength{\tabcolsep}{4pt}
  \caption{Random variables of the running example (slender steel column,
    Euler buckling), with c.o.v.\ denoting the coefficient of variation.}
  \label{tab:running-example-rvs}
  \begin{tabular}{llllrrr}
    \toprule
    Symbol & Description & Distribution & Unit & Mean & Std. dev. & c.o.v. \\
    \midrule
    $E$ & Elastic modulus       & Lognormal & MPa    & $2.00\times10^{5}$   & $1.00\times10^{4}$   & $0.05$ \\
    $I$ & Second moment of area & Lognormal & m$^4$  & $8.33\times10^{-6}$  & $4.00\times10^{-7}$  & $0.05$ \\
    $L$ & Column length         & Normal    & m      & $3.00$               & $0.03$               & $0.01$ \\
    $P$ & Axial load            & Gumbel    & kN     & $1000$               & $200$                & $0.20$ \\
    \bottomrule
  \end{tabular}
\end{table}

\begin{figure}[t]
  \centering
  \includegraphics[width=\linewidth]{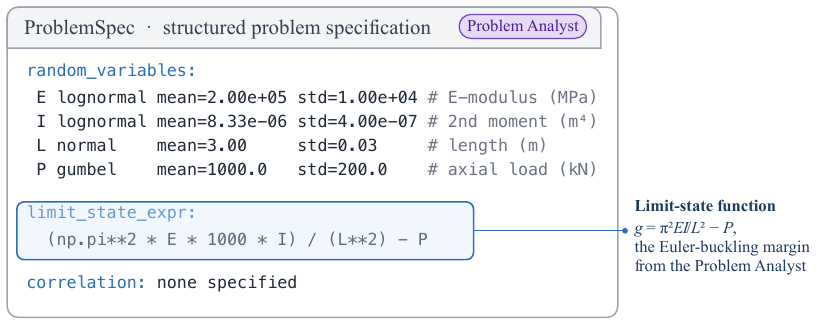}
  \caption{Structured \texttt{ProblemSpec} produced by the Problem Analyst from
    the natural-language input of the running example: four random variables
    with their distributions and parameters, and the limit-state expression.}
  \label{fig:walkthrough-spec}
\end{figure}

The workflow begins with the \emph{Orchestrator}, which classifies the input as a new reliability-analysis request and routes it to the Problem Analyst. The \emph{Problem Analyst} then converts the free-form statement into the structured \texttt{ProblemSpec} defined in Section~\ref{sec:contracts}. For this example, it extracts the four random variables and their distributions, namely \(E\) and \(I\) as lognormal variables, \(L\) as a normal variable, and \(P\) as a Gumbel variable. It also translates the limit-state function into the machine-evaluable expression \texttt{(np.pi**2 * E * 1000 * I) / (L**2) - P} and records the stated independence of the variables as an empty correlation matrix. Because the input already provides all required distributions and parameters, the specification is judged complete and the workflow proceeds without entering the Clarifier loop. Fig.~\ref{fig:walkthrough-spec} shows the resulting \texttt{ProblemSpec}.

\begin{figure}[t]
  \centering
  \includegraphics[width=\linewidth]{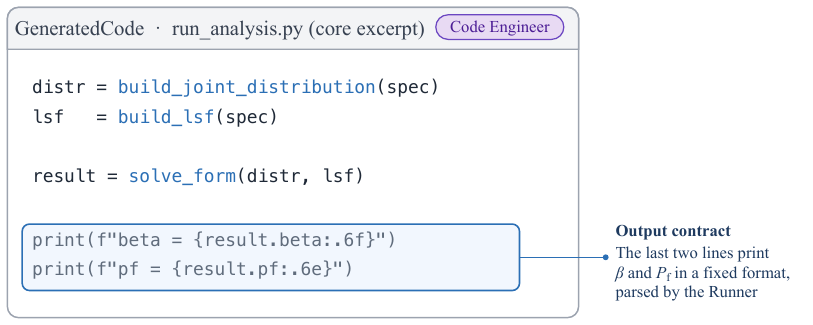}
  \caption{Core of the self-contained Python script generated by the Code
    Engineer for the running example. The specification is turned into a joint
    distribution and a limit-state function, the validated FORM solver is
    invoked, and the results are printed on the last two lines per the output
    format.}
  \label{fig:walkthrough-code}
\end{figure}

Using the \texttt{ProblemSpec}, the \emph{Planner} recommends the FORM-based category. This recommendation is consistent with the problem characteristics: the limit-state function is smooth and differentiable, and the failure probability is not in an extremely rare range, with an expected order of approximately \(10^{-3}\). The recommendation is presented at the user-confirmation gate, where the user accepts the suggested category. The subsequent method-resolution step therefore selects FORM as the concrete method within the FORM-based category. The \emph{Code Engineer} then converts the finalized specification and selected method into a self-contained Python script. The generated code imports the in-house \texttt{reliability\_llm.solvers} package, constructs the joint distribution and limit-state function, calls the FORM solver, and prints the results in the fixed output format required in Section~\ref{sec:agents}. The core of the generated script is shown in Fig.~\ref{fig:walkthrough-code}. The reliability computation is performed by the backend solver, \texttt{solve\_form}, while the Code Engineer translates the structured specification into the corresponding solver call.

The \emph{Runner} executes the generated code in an isolated subprocess and parses the \texttt{beta} and \texttt{pf} values from standard output. For this example, the execution succeeds and yields a reliability index of \(\beta=2.663\) and a failure probability of \(P_f=3.87\times10^{-3}\). Finally, the \emph{Result Interpreter} converts these numerical values into an engineering interpretation following the \texttt{EngineeringInterpretation} data contract in Table~\ref{tab:contracts}. The interpretation includes a concise summary, a reliability rating, a comparison with a target reliability level, and relevant caveats. The generated summary is as follows:

\begin{quote}
``The reliability index $\beta = 2.663$ corresponds to a failure probability
of approximately $3.87\times10^{-3}$, or about 1 in 258. This indicates a
moderate level of reliability.''
\end{quote}

The target comparison indicates that the result is lower than the reference value \(\beta=3.8\) specified in Eurocode EN~1990 for the ultimate limit state over a 50-year reference period. The interpretation also highlights three points relevant to the reliability estimate: the linearization approximation used in FORM, possible physical dependence among the variables despite the stated independence assumption, and the representativeness of the Gumbel tail model for the axial load; Fig.~\ref{fig:gui-result} shows how this interpretation is returned to the user in the interface. Thus, the framework provides both numerical estimates of \(\beta\) and \(P_f\) and their engineering meaning, improving accessibility for non-expert users.

\section{Experiments}
\label{sec:experiments}

This section evaluates the proposed framework on the held-out benchmark described in Section~\ref{sec:benchmark}. Section~\ref{sec:e2e} evaluates the end-to-end workflow, including the execution of the generated code and the numerical accuracy of the resulting reliability estimates, whereas Section~\ref{sec:planner-eval} separately evaluates the method-category selection of the Planner.

\subsection{Benchmark suite}
\label{sec:benchmark}

To evaluate the proposed framework, we construct a held-out benchmark consisting of 20 hand-authored structural reliability problems. These problems are disjoint from the 150 synthetic examples used to train the Planner and are therefore used to assess generalization to unseen reliability-analysis cases. As summarized in Table~\ref{tab:benchmark}, the benchmark covers diverse conditions, including 11 probability distribution types, problem dimensions ranging from 2 to 30 random variables, linear to strongly nonlinear limit-state functions, and failure probabilities ranging from common to rare events. This diversity allows the benchmark to test the framework under conditions where no single reliability method is uniformly suitable. The running example of Section~\ref{sec:running-example} corresponds to problem p04 of this benchmark.

\begin{table}[t]
  \centering
  \caption{Coverage of the held-out benchmark (20 problems).}
  \label{tab:benchmark}
  \begin{tabular}{@{}lp{0.53\columnwidth}@{}}
    \toprule
    Property & Coverage \\
    \midrule
    Number of problems              & 20 (held-out) \\
    Probability distributions       & 11 types (normal, lognormal, Gumbel, Weibull, Fr\'echet, Beta, chi-square, exponential, Rayleigh, Type-I/III smallest) \\
    Dimension (\# random variables) & 2--30 \\
    Limit-state nonlinearity        & linear / mild / strong \\
    Method category (label)         & FORM-based 8 $\cdot$ MCS 6 $\cdot$ Advanced 6 \\
    \bottomrule
  \end{tabular}
\end{table}

The benchmark is used for two evaluations. The first is the end-to-end execution evaluation of the full pipeline in Section~\ref{sec:e2e}. In this evaluation, a run is considered successful when the Code Engineer generates executable code and the Runner returns parseable estimates of \(\beta\) and \(P_f\). The pipeline results are then compared with reference solutions obtained from validated solvers. For all problems except p10 and p17, the reference solution is obtained by MCS; for these two rare-event problems, subset simulation is used instead. The reference analyses and end-to-end evaluations use the same fixed random seed. The resulting end-to-end execution results are summarized in Table~\ref{tab:e2e}, which also reports, for each reference estimate, its estimated coefficient of variation (c.o.v.) and the number of limit-state function evaluations (\(N\)) on which it is based.

In this evaluation, all five LLM-based agents, namely the Orchestrator, Problem Analyst, Method Planner, Code Engineer, and Result Interpreter, are run with a temperature of zero. For the Method Planner, which uses the fine-tuned Gemma~4 E4B model, this setting corresponds to greedy decoding; the other four agents use Gemma~4 26B-A4B with reasoning mode disabled. The output of each LLM invocation is limited to 2,048 tokens. Each invocation generates a single response, and no majority voting or aggregation over multiple candidate responses is used. If a response does not conform to the required output schema, the model is asked to correct it, up to two additional times, and the first valid response is used.

\begin{table}[t]
  \centering
  \scriptsize
  \caption{End-to-end results and reference-estimate precision for the 20 benchmark problems.}
  \label{tab:e2e}
  \setlength{\tabcolsep}{3pt}
  \resizebox{\linewidth}{!}{%
  \begin{tabular}{@{}llrrrrrr@{}}
    \toprule
    & Predicted & \multicolumn{4}{c}{Reference} & \multicolumn{2}{c}{Proposed} \\
    \cmidrule(lr){3-6}\cmidrule(l){7-8}
    Problem & category (method) & $\beta$ & $P_f$ & c.o.v.\ (\%) & $N$ & $\beta$ & $P_f$ \\
    \midrule
    p01~~R--S (normal)                  & FORM-based (FORM) & 4.157 & $1.6\times10^{-5}$ & 3.9 & $4.1\times10^{7}$ & 4.160 & $1.6\times10^{-5}$ \\
    p02~~beam yield                     & FORM-based (FORM) & 1.574 & $5.8\times10^{-2}$ & 4.0 & $1.0\times10^{4}$ & 1.562 & $5.9\times10^{-2}$ \\
    p03~~concrete compression           & FORM-based (FORM) & 4.378 & $6.0\times10^{-6}$ & 4.5 & $8.2\times10^{7}$ & 4.397 & $5.5\times10^{-6}$ \\
    p04~~column buckling (Euler)        & FORM-based (FORM) & 2.649 & $4.0\times10^{-3}$ & 4.1 & $1.5\times10^{5}$ & 2.663 & $3.9\times10^{-3}$ \\
    p05~~exponential ratio              & MCS (MCS)         & 2.733 & $3.1\times10^{-3}$ & 4.6 & $1.5\times10^{5}$ & 2.731 & $3.2\times10^{-3}$ \\
    p06~~trig.\ oscillator              & MCS (MCS)         & 3.585 & $1.7\times10^{-4}$ & 4.8 & $2.55\times10^{6}$ & 3.515 & $2.2\times10^{-4}$ \\
    p07~~fatigue (Weibull)              & MCS (MCS)         & 2.489 & $6.4\times10^{-3}$ & 4.7 & $7.0\times10^{4}$ & 2.489 & $6.4\times10^{-3}$ \\
    p08~~containment (rare)             & FORM-based (FORM) & 3.930 & $4.2\times10^{-5}$ & 4.9 & $1.0\times10^{7}$ & 3.918 & $4.5\times10^{-5}$ \\
    p09~~pile group (corr.)             & MCS (MCS)         & 0.817 & $2.1\times10^{-1}$ & 2.0 & $1.0\times10^{4}$ & 0.813 & $2.1\times10^{-1}$ \\
    p10~~offshore jacket (10 RV)$^{\dagger\ast}$ & FORM-based (FORM) & 5.625 & $9.3\times10^{-9}$ & 4.9 & $3.0\times10^{5}$ & 5.967 & $1.2\times10^{-9}$ \\
    p11~~bounded resistance (Beta)      & FORM-based (FORM) & 2.194 & $1.4\times10^{-2}$ & 4.8 & $3.0\times10^{4}$ & 2.247 & $1.2\times10^{-2}$ \\
    p12~~slope (Mohr--Coulomb)          & MCS (MCS)         & 3.022 & $1.3\times10^{-3}$ & 3.6 & $6.3\times10^{5}$ & 3.011 & $1.3\times10^{-3}$ \\
    p13~~vibration (chi-square)         & MCS (MCS)         & 0.819 & $2.1\times10^{-1}$ & 2.0 & $1.0\times10^{4}$ & 0.794 & $2.1\times10^{-1}$ \\
    p14~~wind extreme (Fr\'echet)       & Advanced (SuS)    & 2.161 & $1.5\times10^{-2}$ & 4.6 & $3.0\times10^{4}$ & 2.097 & $1.8\times10^{-2}$ \\
    p15~~wave overtopping (Rayleigh)    & FORM-based (FORM) & 2.183 & $1.5\times10^{-2}$ & 4.8 & $3.0\times10^{4}$ & 2.178 & $1.5\times10^{-2}$ \\
    p16~~bridge girder (corr.\ loads)$^{\ast}$   & FORM-based (FORM) & 1.998 & $2.3\times10^{-2}$ & 3.8 & $3.0\times10^{4}$ & 3.032 & $1.2\times10^{-3}$ \\
    p17~~parallel system (30 RV)$^\dagger$ & Advanced (SuS)  & 4.702 & $1.3\times10^{-6}$ & 4.2 & $2.1\times10^{5}$ & 4.783 & $8.6\times10^{-7}$ \\
    p18~~yield (Type-I smallest)        & FORM-based (FORM) & 2.178 & $1.5\times10^{-2}$ & 4.7 & $3.0\times10^{4}$ & 2.328 & $1.0\times10^{-2}$ \\
    p19~~shifted (Type-III smallest)    & FORM-based (FORM) & 1.644 & $5.0\times10^{-2}$ & 4.4 & $1.0\times10^{4}$ & 1.633 & $5.1\times10^{-2}$ \\
    p20~~exponential degradation        & FORM-based (FORM) & 2.045 & $2.0\times10^{-2}$ & 4.0 & $3.0\times10^{4}$ & 2.059 & $2.0\times10^{-2}$ \\
    \bottomrule
  \end{tabular}}

  \smallskip
  \parbox{\linewidth}{\raggedright\scriptsize Note: References are MCS results, except SuS for $\dagger$. $\ast$ marks the main discrepancies discussed in Section~\ref{sec:e2e}. This evaluation uses Problem Analyst-generated inputs, whereas the standalone Planner evaluation in Fig.~\ref{fig:confusion} uses predefined inputs.}
\end{table}

The second evaluation concerns the method-selection accuracy of the Planner in Section~\ref{sec:planner-eval}. For this purpose, each benchmark problem is assigned a reference method category using the deterministic labeling criterion described in Section~\ref{sec:finetuning}. Specifically, each problem is solved offline using the reliability-analysis methods considered in this study, and the resulting reference estimates of \(\beta\) and \(P_f\) are used to assign the method-category label. The same labeling rule used to train the Planner is applied to the benchmark problems. Under this rule, the 20 benchmark problems are divided into 8 FORM-based, 6 MCS, and 6 Advanced cases, as also reported in Table~\ref{tab:benchmark}.

\subsection{End-to-end execution}
\label{sec:e2e}

The proposed framework successfully completed the full execution workflow for all 20 benchmark problems. In every case, the Code Engineer generated executable code, and the Runner returned parseable estimates of \(\beta\) and \(P_f\), as reported in Table~\ref{tab:e2e}. For the majority of the benchmark problems, the absolute difference in the reliability index between the pipeline estimate and the reference solution was less than 0.1. This result indicates that the framework can translate natural-language reliability problems into executable analyses and produce numerical reliability estimates consistently.

The main discrepancies occur in p16 and p10, for which the FORM-based category was selected. For p16, the pipeline gives \(\beta=3.03\), compared with the reference value of \(\beta=2.00\). This problem has a strongly nonlinear limit-state function, for which FORM and MCS give substantially different results. For p10, the pipeline gives \(P_f=1.2\times10^{-9}\), compared with the reference value of \(P_f=9.3\times10^{-9}\). These cases indicate room for further improvement in the numerical accuracy of the workflow. As described in Section~\ref{sec:confirmation}, after the first-pass FORM-based analysis, the user is offered a one-time refinement using SORM or FORM-based importance sampling, so that a questionable FORM estimate need not be accepted as the final answer. By contrast, for p14, the framework selected an advanced method for a problem that could also be handled by a simpler method, without a substantial loss of accuracy, although the advanced method typically involves a higher computational cost. Overall, the results show that the framework can complete the end-to-end reliability-analysis workflow without expert intervention, while also indicating that accurate method selection is critical to the quality of the final reliability estimates.

To compare the proposed workflow with direct LLM answers, the same 20 benchmark problems were also presented to Gemma~4 (\texttt{gemma4:26b}), the main model used by the non-Planner agents. Each original natural-language problem statement was followed by the instruction ``Solve this problem and provide the reliability index and the probability of failure.'' A single answer was generated per problem, without a system prompt, tools, code execution, or retries. The temperature was set to zero, reasoning mode was disabled, and the output was limited to 2,048 tokens. All 20 answers were completed without truncation. The final stated values of \(\beta\) and \(P_f\) were extracted and compared with the reference solutions reported in Table~\ref{tab:e2e}.

\begin{figure}[t]
  \centering
  \includegraphics[width=3.9in]{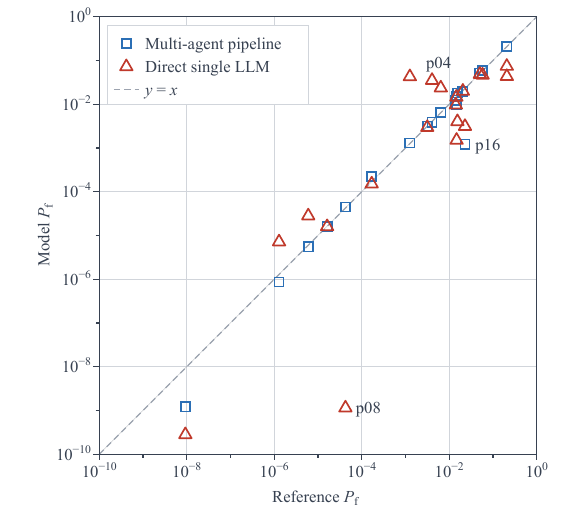}
  \caption{Comparison of failure probabilities from the multi-agent pipeline (blue
    squares) and direct single-LLM answers (red triangles) against the reference
    solutions for the 20 benchmark problems. Both axes are logarithmic, and the
    dashed line denotes exact agreement. Labels identify the direct answers for p04
    and p08 and the pipeline estimate for p16.}
  \label{fig:direct-baseline}
\end{figure}

Fig.~\ref{fig:direct-baseline} shows that the proposed framework agrees substantially more closely with the reference solutions than the direct single-LLM answers. Agreement is quantified by the relative error in \(P_f\), defined as
\begin{equation}\label{eq:pf-relative-error}
\varepsilon_{P_f} = \frac{\lvert P_f - P_{f,\mathrm{ref}} \rvert}{P_{f,\mathrm{ref}}} \times 100\%,
\end{equation}
where \(P_f\) is the failure probability estimated by the framework or stated in the direct answer, and \(P_{f,\mathrm{ref}}\) is the corresponding reference solution. The median relative error is 3.8\% for the framework and 76.6\% for the direct answers. The relative error exceeds 20\% in 5 of the 20 problems for the framework and in 13 for the direct answers, and it exceeds 50\% in 2 and 12 problems, respectively. The direct answers range from close agreement to large discrepancies: for p04, the direct answer overestimates \(P_f\) by a factor of approximately 8.7, whereas for the rare-event problem p08 it underestimates \(P_f\) by more than four orders of magnitude. Although the FORM-based estimate for p16 remains less accurate than the direct answer, the framework achieves substantially lower errors across the benchmark, as reflected in both the median relative error and the number of cases exceeding the error thresholds.

The answer generated by the direct single LLM for p04 also illustrates the difference between a stated numerical answer and an executed analysis. The answer states, ``Running this simulation yields the following approximate values:'', although the model had no access to code execution. These observations support an integrated workflow that generates and executes solver code, rather than relying solely on numerical answers stated by a language model.

\begin{figure}[t]
  \centering
  \includegraphics[width=3.4in]{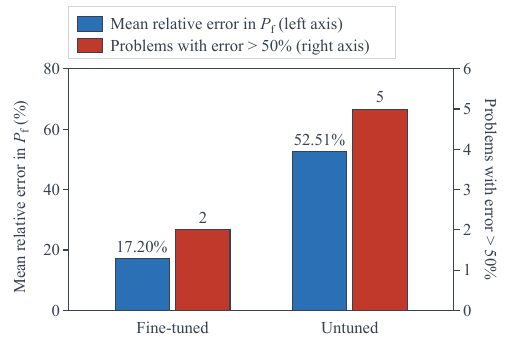}
  \caption{End-to-end results with the fine-tuned and untuned Planners: mean relative error in
    \(P_f\) (left axis) and number of problems with a relative error exceeding 50\% (right axis).
    Both summaries are calculated from the 20 problems executed successfully with the fine-tuned
    Planner and the 19 with the untuned Planner, excluding the execution failure for p07.}
  \label{fig:planner-e2e}
\end{figure}

In addition, Fig.~\ref{fig:planner-e2e} compares the end-to-end numerical errors obtained with the fine-tuned and untuned Planners, providing a separate assessment of the contribution of Planner fine-tuning to the final reliability estimates. For this comparison, the fine-tuned workflow results reported in Table~\ref{tab:e2e} were compared with those obtained using the untuned Gemma~4 E4B Planner on the same 20 benchmark problems, with all other workflow settings unchanged. The workflow returned estimates for all 20 problems with the fine-tuned Planner, whereas it returned estimates for 19 problems with the untuned Planner, failing to return an estimate for p07. The relative-error summaries include the problems successfully completed under each configuration.

For p14, the FORM-based estimate obtained with the untuned Planner was closer to the reference solution than the estimate from the simulation-based method selected by the fine-tuned Planner. For other problems, however, the FORM-based category selected by the untuned Planner led to large errors: for p12 and p17, the relative errors were 311\% and 319\%. Over the evaluated benchmark, the fine-tuned Planner yielded a lower mean relative error and fewer problems with relative errors exceeding 50\%.

\subsection{Evaluation of the Planner}
\label{sec:planner-eval}

This section evaluates the Planner independently from the rest of the pipeline. Each model receives the same prompt and the same \texttt{ProblemSpec} for each held-out benchmark problem, and its predicted method category is compared with the reference label. Category accuracy measures agreement between the Planner's predictions and the reference labels assigned by the deterministic labeling criterion. The numerical accuracy of the final reliability estimates is evaluated separately for the end-to-end workflow in Section~\ref{sec:e2e}. 

Three open-weight configurations are compared: (i) Gemma~4 E4B without fine-tuning, using prompting only; (ii) Gemma~4 26B-A4B without fine-tuning, which is the backbone used for the non-Planner agents in the deployed framework; and (iii) Gemma~4 E4B with Planner-only QLoRA fine-tuning, which is the proposed Planner. The prompt is identical across the three configurations and is reproduced in \ref{app:planner-prompt}. In the main evaluation, all three configurations use greedy decoding, and a single response is generated for each problem. The evaluation metric is category accuracy, reported both overall and separately for the FORM-based, MCS, and Advanced categories. 

\begin{figure}[t]
  \centering
  \includegraphics[width=\linewidth]{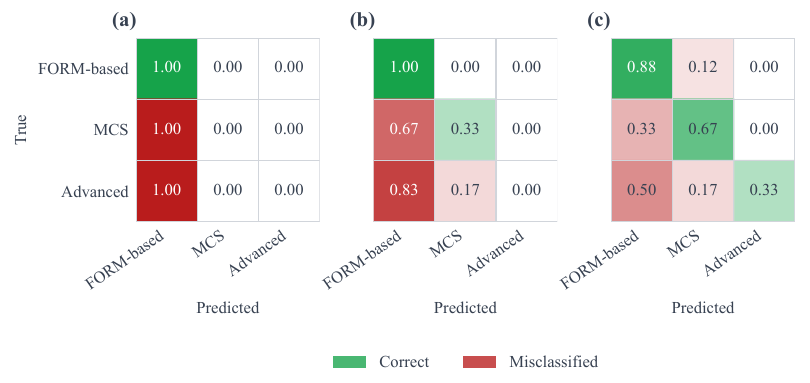}
  \caption{Confusion matrices of method-category selection on the held-out
    benchmark ($n=20$; class sizes: FORM-based 8, MCS 6, Advanced 6). Each row is
    normalized to sum to one, so the diagonal entries are the per-class recall.
    Green marks correct predictions (the diagonal) and red marks misclassifications
    (off-diagonal), with color saturation proportional to the fraction.
    (a)~Gemma~4 E4B without fine-tuning (overall 40\%); (b)~Gemma~4 26B-A4B without
    fine-tuning (50\%); (c)~Gemma~4 E4B with Planner-only QLoRA (ours, 65\%). The
    two models without fine-tuning, (a) and (b), collapse onto the FORM-based
    column---predicting FORM-based even for the MCS and Advanced problems---whereas
    the fine-tuned model (c) recovers the minority categories.}
  \label{fig:confusion}
\end{figure}

Fig.~\ref{fig:confusion} shows the category-selection results as confusion matrices. Even though the decision criteria are provided in the prompt, the two untuned models default to the FORM-based category for most problems: they identify all 8 FORM-based cases but none (E4B) or 2 (26B-A4B) of the 6 MCS cases and none of the 6 Advanced cases. This bias is the unfavorable direction of misselection: because FORM relies on a first-order approximation of the limit-state surface, applying it to strongly nonlinear or rare-event problems can yield a numerically valid but biased estimate. In contrast, the fine-tuned Planner correctly identifies 7 of 8 FORM-based, 4 of 6 MCS, and 2 of 6 Advanced cases, extending correct selections beyond the FORM-based category. Its overall accuracy is 65\%, compared with 40\% for the untuned E4B model and 50\% for the larger untuned 26B-A4B model. In addition, routing a FORM-based problem to MCS or Advanced is not necessarily an unsuitable choice merely because the category differs from the reference label, since these simulation-based methods can also be applied to such problems, typically at a higher computational cost. We therefore also report a supplementary acceptable-selection rate defined for this study: for problems labeled FORM-based, any of the three categories is counted as acceptable, whereas for problems labeled MCS or Advanced, only the labeled category is acceptable, and MCS and Advanced are not accepted in place of each other. Under this criterion, the fine-tuned Planner makes an acceptable selection for 14 of the 20 problems (70\%), including one FORM-labeled problem routed to MCS, whereas the untuned models remain at 40\% (E4B) and 50\% (26B-A4B) because all of their misselections involve MCS- or Advanced-labeled problems. This rate complements category accuracy and does not measure numerical accuracy or computational efficiency. These results support specializing the Planner within the workflow: on this benchmark, fine-tuning reduces the bias toward the FORM-based category and recovers the minority MCS and Advanced categories.

To examine the sensitivity of these results to stochastic decoding, the untuned and fine-tuned E4B models were additionally evaluated by sampling at temperatures of 0.3 and 0.7, with five generation seeds (0--4) for each model and temperature. Each combination of model, temperature, and seed was treated as an independent run, in which a single response was sampled for each problem; the metrics were computed for each run, without majority voting across runs. This comparison reused the 20 benchmark problems, the prompt, and the E4B checkpoints of the main evaluation. Both models were run with the same inference backend, without retraining. The maximum output length was 32 tokens, and the top-k and top-p values were 64 and 0.95, respectively.

Table~\ref{tab:planner-sampling} summarizes the results. Between-seed variation in category accuracy occurred only for the fine-tuned Planner at a temperature of 0.7, whereas the acceptable-selection rate remained unchanged across seeds for both models at both temperatures. Nevertheless, the fine-tuned Planner achieved higher values for both metrics than the untuned E4B model in every sampled run at both temperatures.

\begin{table}[t]
  \centering
  \caption{Category accuracy and acceptable-selection rate of the E4B Planner under stochastic sampling on the 20 held-out benchmark problems. Values are means $\pm$ sample standard deviations across five generation seeds; means are given in percent and standard deviations in percentage points.}
  \label{tab:planner-sampling}
  \small
  \begin{tabular}{@{}lccc@{}}
    \toprule
    Model & Temperature & Category & Acceptable-selection \\
          &             & accuracy (\%) & rate (\%) \\
    \midrule
    \begin{tabular}[c]{@{}l@{}}Gemma~4 E4B\\without fine-tuning\end{tabular}     & 0.3 & $40.0 \pm 0.0$ & $40.0 \pm 0.0$ \\
    \begin{tabular}[c]{@{}l@{}}Gemma~4 E4B\\without fine-tuning\end{tabular}     & 0.7 & $40.0 \pm 0.0$ & $40.0 \pm 0.0$ \\
    \begin{tabular}[c]{@{}l@{}}Gemma~4 E4B with\\Planner-only QLoRA\end{tabular} & 0.3 & $65.0 \pm 0.0$ & $70.0 \pm 0.0$ \\
    \begin{tabular}[c]{@{}l@{}}Gemma~4 E4B with\\Planner-only QLoRA\end{tabular} & 0.7 & $64.0 \pm 2.2$ & $70.0 \pm 0.0$ \\
    \bottomrule
  \end{tabular}
\end{table}

\section{Discussion}
\label{sec:discussion}

\subsection{Accessibility and trustworthiness of the framework}
\label{sec:accessibility-discussion}

The main significance of the proposed framework is that it lowers the expertise barrier of structural reliability analysis while preserving computational trustworthiness. The framework enables users to proceed from a natural-language problem statement to interpreted estimates of \(\beta\) and \(P_f\). However, these numerical quantities are not generated by an LLM. They are obtained through deterministic execution of validated reliability solvers, which reduces the risk of hallucinated numerical results and makes the reported estimates reproducible. The comparison with direct single-LLM answers in Section~\ref{sec:e2e} illustrates the importance of this design. For p04, the direct answer states, ``Running this simulation yields the following approximate values,'' despite having no access to code execution, as discussed in Section~\ref{sec:e2e}. In contrast, the framework obtains its reliability estimates from executed solver code.

Trustworthiness is further supported by the structured workflow. Because intermediate outputs are stored through data contracts, the final reliability estimates can be traced back to the problem specification, selected method, generated code, and execution result. These intermediate outputs are also exposed in the trace panel of each system turn of the web interface (Section~\ref{sec:ui} and Fig.~\ref{fig:gui}(b)), so that the user can inspect how the reported estimates were obtained. This traceability is important in design-verification contexts, where reliability assessments should be reproducible and defensible rather than presented as opaque model outputs. 

Accessibility is supported by the human-in-the-loop design. The framework pauses at key decision points, allowing users to confirm, revise, or ask questions before the analysis proceeds. This allows non-expert users to retain control without requiring full knowledge of reliability theory, solver implementation, or method-specific assumptions. These interactions occur through the chat-based web interface. Expert users may also benefit because the framework reduces repetitive tasks such as problem encoding, solver execution, and result reporting.

Local execution is a further practical feature of the framework. Because all agents use open-weight models, they can run locally together with the Runner, the backend solvers, and the web interface (Section~\ref{sec:ui}), without dependence on closed APIs. Consequently, the problem statement, intermediate outputs, and results need not be sent to an external model provider. This design supports reproducibility, data control, and adoption in engineering settings where external data transfer or proprietary model access may be undesirable.

\subsection{Domain-specialized method selection via Planner-only fine-tuning}
\label{sec:why-planner-only}

On the held-out benchmark, Planner-only fine-tuning improves method-category selection relative to both untuned configurations. The fine-tuned Gemma~4 E4B Planner reaches an exact category accuracy of 65\%, compared with 40\% for the untuned E4B model and 50\% for the larger untuned Gemma~4 26B-A4B model, and an acceptable-selection rate of 70\%, compared with 40\% and 50\%, under the supplementary criterion defined in Section~\ref{sec:planner-eval}. This matters because the selected category determines the solver that computes the reliability estimate. The untuned models tend to over-predict the FORM-based category, even when the prompt provides the decision criteria. Such a bias is problematic because FORM is efficient but can be inaccurate for strongly nonlinear limit-state functions or rare-event problems. In these cases, selecting FORM-based methods by default can produce an apparently valid but unreliable estimate. On this benchmark, Planner-only fine-tuning reduces this tendency by improving discrimination among the FORM-based, MCS, and Advanced categories. On the 20-problem held-out benchmark, the fine-tuned E4B Planner achieved higher category accuracy and acceptable-selection rate than the larger untuned 26B-A4B model. Further comparisons with state-of-the-art closed-source LLMs are needed to assess whether this performance advantage extends beyond the evaluated open-weight configurations. Method-category selection is an \emph{a priori} decision: the numerical results that indicate whether a method is suitable, including those used to assign the reference labels, become available only after the analysis, so the suitable method generally cannot be established with certainty beforehand. The Planner therefore recommends an initial method category from the \texttt{ProblemSpec}, and fine-tuning improves this \emph{a priori} recommendation. To account for the uncertainty inherent in such a prior choice, the workflow is designed such that the user can confirm the recommendation or switch to another category before code generation and numerical execution begin (Section~\ref{sec:confirmation}). The framework is therefore intended to support, rather than replace, engineering judgment. 

The selective fine-tuning strategy also has practical advantages. Only the Planner is fine-tuned, as a component-level specialization for \emph{a priori} method-category selection, a closed-set decision whose training labels can be generated offline from reference solutions. The remaining agents handle language understanding, structured extraction, code generation, and result interpretation using task-specific prompts. This keeps the adaptation localized and feasible with modest hardware: the QLoRA adapter is trained on a single 24~GB GPU, without full-model retraining or fine-tuning of the other agents. The adapted Planner also remains an open-weight model that runs locally together with the other agents, so the specialization does not introduce a dependence on closed APIs. 

The learned method-selection behavior is also configurable. The category boundaries are determined by the expert-defined labeling criterion used to generate the training data, rather than being fixed properties of the model. For example, a more conservative practitioner could require FORM and MCS estimates to agree within 3\% rather than the 5\% adopted in this study, thereby routing more borderline cases to MCS. Such a change only requires regenerating the labels and retraining the Planner, while leaving the remaining agents unchanged.

\subsection{Limitations and future work}
\label{sec:limitations}

In Section~\ref{sec:experiments}, the framework is evaluated on 20 held-out, hand-authored benchmark problems, which limits the statistical precision of the Planner's reported category accuracy and acceptable-selection rate, as well as the extent to which the results can be generalized beyond these cases. We applied two-sided exact McNemar tests to the paired per-problem correctness of the method-category predictions from the main greedy-decoding evaluation on the 20 benchmark problems. The category-accuracy differences were not statistically significant at the 0.05 level for the fine-tuned E4B Planner compared with either the untuned E4B model (\(p = 0.125\)) or the untuned Gemma~4 26B-A4B model (\(p = 0.375\)). The higher accuracy of the fine-tuned Planner should therefore be interpreted as an observed improvement within the evaluated cases, rather than statistically established superiority over the untuned configurations. The Planner's exact category accuracy of 65\% also leaves room for further improvement. The observed improvements should also be interpreted as findings on this benchmark rather than evidence of broad generalization or robustness across structural reliability problems. The Planner comparison focused on open-weight LLM configurations selected for local execution without dependence on closed APIs. Because frontier LLMs and classical classifiers were not evaluated, the present results do not establish an advantage over these alternatives or demonstrate that LLM fine-tuning is necessary for this task. The comparative cost-effectiveness of Planner fine-tuning was not assessed in this study. Such an assessment would require a separate comparison of training and deployment costs. The supplementary evaluation in Section~\ref{sec:planner-eval} quantifies the variability of the Planner's selections under stochastic decoding. The variability of the complete end-to-end workflow was not assessed. In addition, the present framework is limited to component-level structural reliability analysis. System configurations, such as series and parallel systems, can be represented by combining component limit-state functions using \(\min\) or \(\max\), but the framework does not yet include dedicated system-reliability algorithms. Direct simulation remains conceptually valid, but it may become computationally expensive when the system failure probability is small. The framework is also currently limited to limit-state functions that can be evaluated directly within the generated code. Problems requiring external numerical models, such as finite-element analysis, are not considered in the present implementation. Coupling the framework with structural analysis solvers such as OpenSeesPy~\citep{zhu2018openseespy} would extend its applicability to more realistic engineering problems. 

In the proposed workflow, the final reliability estimates can be traced back to the problem specification, selected method, generated code, and execution result (Section~\ref{sec:accessibility-discussion}), and the Runner returns structured error information, including the failure type, when execution fails (Section~\ref{sec:agents}). This traceability supports inspection of each analysis but does not by itself ensure numerical accuracy, which is evaluated against the reference solutions in Section~\ref{sec:e2e}. The one-time refinement described in Section~\ref{sec:confirmation} was not exercised in the benchmark evaluation, so its effectiveness was not assessed in this study.

Several extensions follow from these limitations. First, incorporating efficient system-reliability methods, such as probability bounds and cut-set/path-set formulations~\citep{song2021structural}, would improve tractability for general system configurations. Second, linking the framework with finite-element solvers would allow reliability assessment of problems whose limit states are not available in closed form. Third, accepting multimodal inputs, such as drawings or structural models, and expanding the benchmark toward real-world engineering cases would provide a broader test of practical applicability. Finally, increasing the Planner's training data and expanding the set of selectable reliability methods may further improve method-selection accuracy. 

Another important extension is to ground standards-related information in authoritative documents. In the present framework, the user provides the probabilistic model, including distribution types and parameters, while the target reliability level is handled at the interpretation stage using preset reference values and model judgment. This information is not yet retrieved directly from governing codes. RAG offers a natural way to address this issue. For example, distribution models for loads and material properties could be retrieved from sources such as the JCSS Probabilistic Model Code~\citep{jcss2001pmc} or the Eurocodes~\citep{cen2002en1990}. Similarly, target reliability levels, such as the target \(\beta\) values associated with EN~1990 reliability classes, could be retrieved from the relevant code documents and used during result interpretation. Because adapting to another code would require updating the retrieval corpus rather than retraining the model, this approach could support project-specific and jurisdiction-specific reliability assessment. RAG has also shown promise in document-based construction-engineering tasks~\citep{uhm2025effectiveness}.

\section{Conclusion}
\label{sec:conclusion}

This study presented a multi-agent large language model (LLM) framework designed for component-level structural reliability analysis, specifically the estimation and interpretation of the reliability index \(\beta\) and failure probability \(P_f\). The proposed framework enables users to proceed from a natural-language problem statement to an interpreted reliability result by coordinating specialized agents for problem formulation, method planning, code generation, deterministic execution, and result interpretation. A main feature of the framework is the separation between language-based reasoning and numerical computation: the LLM agents structure the problem, select the analysis category, generate solver code, and explain the result, whereas \(\beta\) and \(P_f\) are computed only through deterministic execution of validated reliability solvers. This design reduces the risk of hallucinated numerical results and supports traceability through structured data contracts. The framework is also built on open-weight models and can run locally, avoiding dependence on closed APIs. In addition, Planner-only QLoRA fine-tuning improved method-category selection accuracy on the 20-problem held-out benchmark. The compact Gemma~4 E4B Planner achieved 65\% accuracy, compared with 40\% for its untuned counterpart and 50\% for the larger untuned Gemma~4 26B-A4B model. These results support the value of targeted domain adaptation for \emph{a priori} method-category selection, including its ability to improve performance relative to the larger untuned model evaluated here. Future comparisons with state-of-the-art closed-source LLMs and classical classifiers would further clarify the comparative merits of this approach. The framework therefore provides a practical step toward making reliability-based assessment more accessible, reproducible, and interpretable for a broader range of engineering users.

\section*{CRediT authorship contribution statement}

\textbf{Jaehwan Jeon}: Conceptualization, Methodology, Software, Investigation,
Validation, Visualization, Writing -- original draft.
\textbf{Chang Hee Lee}: Software, Investigation, Validation, Visualization.
\textbf{Jinhyun Choo}: Conceptualization, Writing -- review \& editing.
\textbf{Taeyong Kim}: Conceptualization, Methodology, Supervision,
Writing -- review \& editing, Project administration, Funding acquisition.

\section*{Declaration of competing interest}

The authors declare that they have no known competing financial interests
or personal relationships that could have appeared to influence the work
reported in this paper.

\appendix

\section{Method Planner system prompt}
\label{app:planner-prompt}

The complete system prompt of the Method Planner is reproduced below. It
states the category-decision criteria used to assign the reference labels
(Section~\ref{sec:benchmark}). All three
configurations compared in Section~\ref{sec:planner-eval} receive this same
prompt.

\begin{lstlisting}[language={},keywordstyle={},numbers=none,
  basicstyle=\ttfamily\footnotesize,
  literate={β}{{$\beta$}}1 {Δ}{{$\Delta$}}1 {≲}{{$\lesssim$}}1
           {⟹}{{$\Rightarrow$}}1 {·}{{$\cdot$}}1 {—}{{---}}1 {…}{{...}}1
           {²}{{$^{2}$}}1 {³}{{$^{3}$}}1 {⁴}{{$^{4}$}}1 {⁶}{{$^{6}$}}1
           {⁺}{{$^{+}$}}1 {⁻}{{$^{-}$}}1 {→}{{$\rightarrow$}}1 {−}{{$-$}}1
           {√}{{$\surd$}}1 {≈}{{$\approx$}}1 {≤}{{$\leq$}}1]
# Method Planner Agent

You pick **one** analysis category for a given `ProblemSpec`. The three
categories map to the practitioner's real decision, which turns on two things:
**how rare is failure** (the reliability index β, equivalently Pf), and **can
FORM's single design-point linearization be trusted?** The runtime cascade then
turns the chosen category into a concrete solver.

## Your output

Return a JSON object matching `AnalysisPlan`:
```json
{ "recommended_category": "FORM-based" | "MCS" | "Advanced" }
```
You are a **pure classifier**: emit ONLY the category — no reasoning, no
confidence field (the user-facing "why" is generated downstream). Pick exactly
one; your answer is scored by exact match to the gold category.

## The decision — apply in THIS order

**1. Is failure RARE?  β > 3.5  (Pf ≲ 2·10⁻⁴)  ⟹ `Advanced`.**
Members: CE, iCE, SIS, SuS, CLS. In the rare-event regime plain MCS is
infeasible (~10⁶⁺ samples to converge) and FORM cannot be trusted as the sole
answer, so route to advanced sampling. **Rarity is THE trigger for Advanced, and
it OVERRIDES everything below** — a strongly curved or multimodal problem that is
ALSO rare is `Advanced`, not `MCS`.

**2. Otherwise (β ≤ 3.5, non-rare) — is FORM accurate?  ⟹ `FORM-based`.**
Members: FORM, SORM, IS. Recommend FORM-based when the limit state is linear or
only mildly nonlinear, the failure region sits near a single design point, and
FORM's linearization would land within a few % of the true (MCS) β. This is the
common, easy case.

**3. Otherwise (β ≤ 3.5 but FORM is UNRELIABLE)  ⟹ `MCS`.**
Member: MCS. When failure is NOT rare but the surface is strongly curved or
multimodal so a single linearization is wrong, you do not need a rare-event
method — just sample directly with MCS for the unbiased reference. **This is the
case the rule exists for: FORM-breaking structure + non-rare = `MCS`, NOT
`Advanced`.** (Curvature alone does not make a problem Advanced; rarity does.)

## What breaks FORM (this is what separates 2 → 3, i.e. FORM-based vs MCS)

FORM linearizes at one design point. It is INACCURATE when the failure surface
is curved or has more than one relevant region. Treat these as FORM-breaking
(⟹ MCS when non-rare, ⟹ Advanced when rare):
- two-sided / band / `abs(...)` limit states (failure on both sides) — bimodal
- `min(...)` / `max(...)` of competing failure modes — kinked surface
- a denominator that blows up near a pole: resonance `1/((1−r²)…)`, P-Δ
  `M/(1−P/Pcr)`, cable sag, eccentric footing `Q/(B−2e)²`
- oscillatory `sin` / `cos` terms — multiple failure lobes
- `√` of a quadratic form / vector magnitude `√(x²+y²)` interaction surfaces
- products or strong curvature where the design point is not representative

NOT FORM-breaking — these stay `FORM-based` when mild and non-rare: a single
monotonic power, a single product, a mild ratio. FORM finds the design point
fine; a monotonic transform does not fool it.
\end{lstlisting}

\section*{Data availability}
The source code, fine-tuned model weights, and benchmark problems will be made
publicly available on GitHub upon acceptance.
Repository: \url{https://github.com/JaehwanJeon/Reliability-LLM-supplementary-material}.

\section*{Acknowledgements}
This research was supported by the Yonsei University Research Fund of 2026-22-0244, the National Research Foundation of Korea (NRF) grant funded by the Korean government (MSIT) (No. RS-2023-00209799), and the Creative-Pioneering Researchers Program through Seoul National University.

\bibliographystyle{elsarticle-num-names}
\bibliography{refs}

\end{document}